\documentclass{article}

\usepackage{arxiv}

\usepackage[utf8]{inputenc} 
\usepackage[T1]{fontenc}    
\usepackage{hyperref}       
\usepackage{url}            
\usepackage{booktabs}       
\usepackage{amsfonts,amsmath}       
\usepackage{nicefrac}       
\usepackage{microtype}      
\usepackage{multirow}
\usepackage{graphicx}
\usepackage{lipsum}
\usepackage{xcolor}
\usepackage{comment}
\usepackage{supertabular,booktabs}

\newcommand{\JB}[1]{\textcolor{magenta}{JB: #1}}

\begin{document}

\title{Risk Management Framework for Machine Learning Security}

\date{}

\author{
Jakub Breier\thanks{This work was partially done when the author was with HP-NTU Digital Manufacturing Corporate Lab, Singapore} \\
Silicon Austria Labs, TU-Graz SAL DES Lab\\
Graz University of Technology\\
Graz, Austria\\
\texttt{jbreier@jbreier.com} \\
\And
Adrian Baldwin \\
HP Labs \\
Bristol, United Kingdom \\
\texttt{adrian.baldwin@hp.com} \\
\And
Helen Balinsky \\
HP Labs \\
Bristol, United Kingdom \\
\texttt{helen.balinsky@hp.com} \\
\And
Yang Liu \\
HP-NTU Digital Manufacturing Corporate Lab \\
Singapore \\
\texttt{yangliu@ntu.edu.sg}
}

\maketitle

\begin{abstract}
Adversarial attacks for machine learning models have become a highly studied topic both in academia and industry. 
These attacks, along with traditional security threats, can compromise confidentiality, integrity, and availability of organization's assets that are dependent on the usage of machine learning models.
While it is not easy to predict the types of new attacks that might be developed over time, it is possible to evaluate the risks connected to using machine learning models and design measures that help in minimizing these risks.

In this paper, we outline a novel framework to guide the risk management process for organizations reliant on machine learning models.
First, we define sets of evaluation factors (EFs) in the data domain, model domain, and security controls domain.
We develop a method that takes the asset and task importance, sets the weights of EFs' contribution to confidentiality, integrity, and availability, and based on implementation scores of EFs, it determines the overall security state in the organization.
Based on this information, it is possible to identify weak links in the implemented security measures and find out which measures might be missing completely.
We believe our framework can help in addressing the security issues related to usage of machine learning models in organizations and guide them in focusing on the adequate security measures to protect their assets.
\end{abstract}



\section{Introduction}
Machine learning (ML) techniques have found their use in many domains of our everyday life, ranging from simple tasks such as search query completion to more advanced such as autonomous driving and automated manufacturing.
As in any area of human development, there are adversaries exploiting the security loopholes, for example for financial gain, fame or research. Organizations commonly use risk assessment frameworks to understand where to place security investments to minimize impact of attacks or loss.

However, existing risk frameworks do not accommodate/extend naturally for unique security vulnerabilities introduced by ML.
Therefore, it is essential to develop additional  methodologies that guide organizations and their security teams in assessing and minimizing the risks associated with attacks on deployed ML solutions. As the ML solutions become used in business critical solutions, including automated processes, the need for such methodologies increases.

After adversarial attacks were shown to be a realistic threat to ML models in 2013 by Szegedy et al.~\cite{szegedy2013intriguing}, a great research effort has been invested into finding new attacks~\cite{goodfellow2014explaining,nguyen2015deep,moosavi2016deepfool} and defenses~\cite{papernot2016distillation,yuan2019adversarial,zantedeschi2017efficient,lee2018simple}.
There are often trade-offs between the performance of an ML model and its robustness to adversarial attacks.
Some systematization efforts were done to navigate through the vast amount of published works (e.g.,~\cite{biggio2018wild,zhang2019machine}), and British Information Commissioner's Office is currently working on a guidance on artificial intelligence (AI) auditing framework~\cite{ico}.
However, a comprehensive approach to risk management is missing.

\begin{figure}
    \centering
    \includegraphics[width=0.55\textwidth]{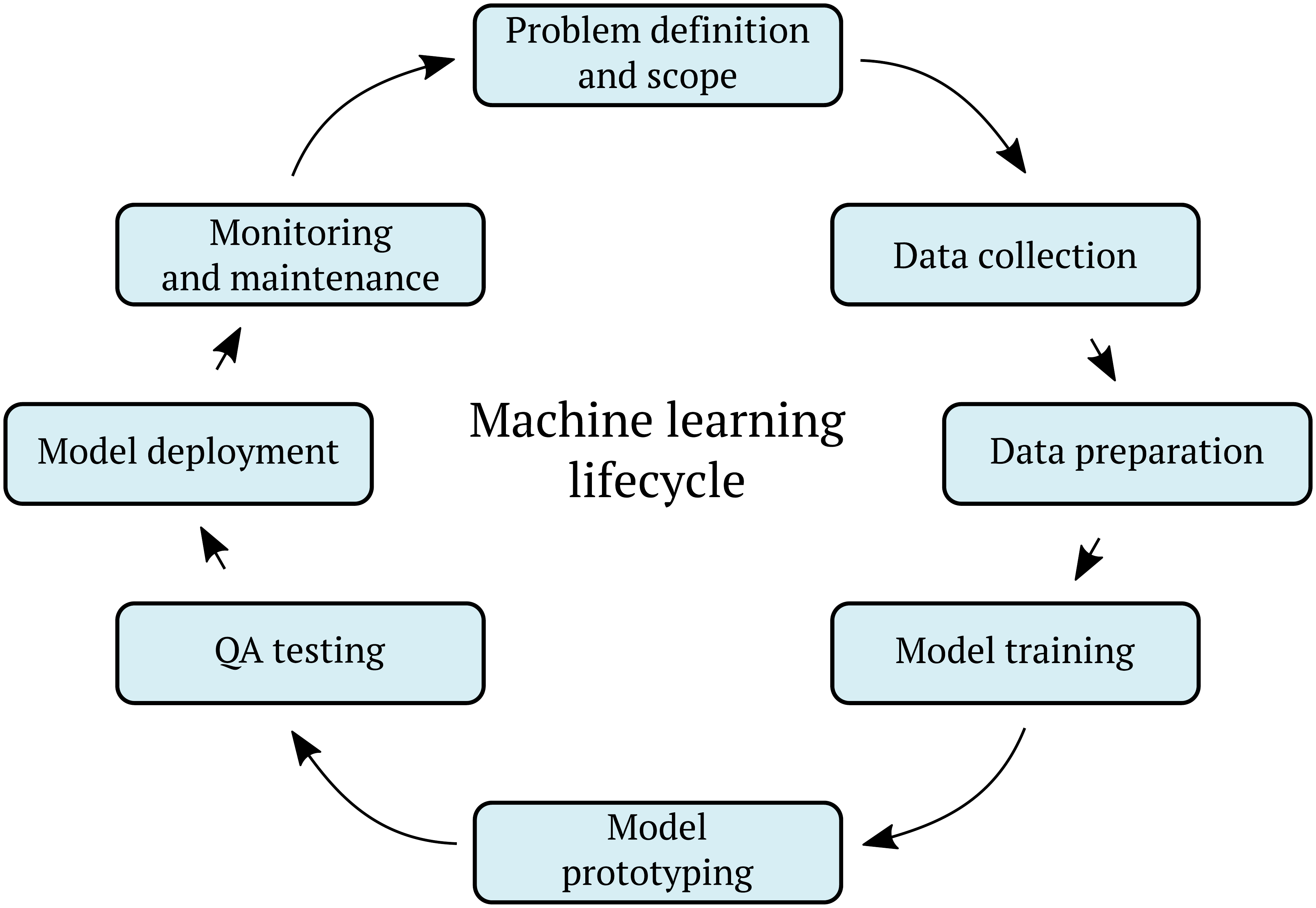}
    \caption{Machine learning lifecycle diagram.}
    \label{fig:lifecycle}
\end{figure}

When considering malicious adversary targeting an ML model, it is useful to employ risk management techniques to find out following:
\begin{itemize}
    \item {\bf Attacker's motivation}: financial gains, industrial espionage and competitive advantage, sabotage of critical infrastructure, personal revenge, outrage or fame.
    \item {\bf Attacker's goal}: theft of confidential information or resources, vandalism, sabotage.
    \item {\bf Attacker's skill and equipment}: recreational attackers with limited technical resources, emotionally committed hacktivists pursuing specific attacks, organized crime with significant technical resources and, finally, highly sophisticated state-sponsored attackers with nearly unlimited resources.
    \item {\bf Attack impact}: loss of life, damage to infrastructure, financial and reputational loss, regulatory fines, intense media scrutiny. 
    \item {\bf Attack cost\footnote{We would like to note that the difference between the attack impact and the attack cost is that the former is the immediate impact after the attack, while the latter indicates the longer term costs after the attack had occured.}}: cost to repair IT infrastructure, consequential cost leading from loss of revenue due to downtime, missed business opportunities, long-term customer attrition, company market value drop, fees and fines from lawsuits and regulatory bodies.
\end{itemize}
Currently, in the ML community, lowering the risk almost exclusively means building more robust models and, for example, reducing misclassifications. 
However, when it comes to risk management approaches, there are high-level techniques that can be used, such as risk mitigation, risk avoidance, transfer of risk, and risk acceptance and these techniques often use a mixed approach assessing people, process and technology. 
Zooming into ML specific methods, practices such as good data hygiene, good model development processes, and good model security help significantly to protect against variety of threats.
The main advantage of using higher level risk management is proactive defense -- in case there is a new attack that was not expected and can affect models that are considered robust against other attacks, risk can be managed in other than technical way to minimize the impact and ensure the business continuity.
Risk management should be continuously applied during the entire ML lifecycle, depicted in Figure~\ref{fig:lifecycle}.

To be able to properly identify which technique is the most appropriate for the given situation, it is necessary to do risk analysis.
It is possible to follow established methodologies, such as one stated in the ISO/IEC 27005~\cite{iso27005} standard to make the entire process repeatable and well-defined.

When looking into identifying security concerns of an ML-enabled system, there are some general questions related to ML model development lifecycle as shown in Fig. \ref{fig:lifecycle} Example questions:
\begin{itemize}
    \item Data collection and preparation
         \subitem -- Were data for ML development collected from trustworthy sources?
         \subitem -- Were appropriate protections placed on the data during transfer and at rest?
         \subitem -- Were integrity checks on datasets performed?
         \subitem --  Were only authorized personnel provided with access to data?
     \item ML model training, fine-tuning and testing      
        \subitem -- Were authentic and non-compromised datasets with good provenance used for training and testing?
        \subitem -- Was it a secure isolated environment with strict access control? 
        \subitem -- Was a trained ML model securely stored and protected from unauthorized copying or modifications?
        \subitem (for supervised learning only) were authorised personnel used for data labelling?
    \item ML model in deployment (inference)
    \subitem -- How a deployed model gets its input?Are there any checks applied to validate the authenticity and integrity of data used for inference?
    \subitem -- How a deployed model is protected from unauthorised access (either direct or indirect for inference) and modifications? 
    \subitem -- Were appropriate protections placed on the data during transfer, if inference performed in cloud?
    \subitem How is the output protected?
    \subitem -- Were appropriate audit logs taken in case decisions are challenged or need review?
    \subitem -- Where multiple models, or redundant (load balanced) solutions are used how are these handled?
    \subitem -- Are the model outputs kept confidential?
    \subitem -- Is any additional validation applied to the output? 
\end{itemize}

Using questions such as these along with knowledge of attacks, the criticality of the tasks, impacts of failures, it is possible to estimate risks associated with a model deployment.
We argue that the risks that might come up with the utilization of ML models differ from risks that are associated with traditional software engineering techniques.
This is due to: data-driven development process (rather than specification-driven) used to construct an ML model which results in an ML model defined by the training data;  substantial difference in development cycles between ML systems and traditional software; the non-deterministic nature of ML systems vs traditional software; large input vectors into ML systems and complex non-linear relationships making it hard-to-impossible to perform exhaustive testing which is very often used in traditional software development. As a result new classes of attacks are constantly emerging.
Same argument applies to mitigation techniques.
 
\medskip

\textbf{Our contribution.} When looking into security requirements of ML models, apart from previous aspects, it is necessary to consider the statistical character of these models -- the probability of getting the correct result might be very high, but it will never reach 1. The profitability of the business hinges on the  proportion of false positive and false negatives vs true positives and associated costs of failures and successes are reflected in the business process.
Adversarial attacks might indirectly try to lower the success probability without being noticed.
As there are many businesses deploying machine learning models nowadays, it is necessary to adjust the risk management methodology to find the related risks and mitigate them.

In this paper, we present a risk management framework for organizations using ML models in their business processes. Our proposed framework is heavily influenced by traditional enterprise risk management frameworks but targeted at the ML specific issues.
We select security controls that provide protection against attacks during ML lifecycle and provide evaluation factors that help in determining the level of security in the organization.
We believe our method can help in designing more secure industrial processes that rely heavily on use of ML.

\section{Related Work}
In this section we provide the necessary background on risk assessment frameworks, adversarial attacks, threat modelling for ML, and multiple-criteria decision making.

\subsection{Risk Assessment Frameworks}
\label{sec:infosec}
Information security aims at lowering cyber security risks.
Cyber security risks are defined as operational risks to information and technology
assets that might affect the \textit{confidentiality}, \textit{availability}, or \textit{integrity} (CIA triad) of information or information systems. 
Standards like ISO 27001~\cite{iso27001} or NIST SP 800-53~\cite{nist800-53} aim at defining risk mitigation techniques to handle the information security risks posed to organizations.
A general cybersecurity framework was recently published by NIST~\cite{nist2018}.
These and many other guidelines provide high-level overview of how the security should be implemented within organizations such that business impact is minimized when security incidents occur.
When using such frameworks it is necessary to check the implementation details of each deployed security control to understand the overall state of the security.
This applies to technical, policy, legal, people and policy related controls.
Barnard and Von Solms provide a good checklist on what security aspects should be evaluated when looking into security controls and mechanisms~\cite{barnard2000formalized}:
\begin{itemize}
    \item Functionality -- indicates whether the proposed controls are present and functional within the organization.
    \item Assurance of correctness -- ensures that the controls are operational and implemented without errors.
    \item Assurance of effectiveness -- ensures that the operation of the controls is consistent with the goals stated in the security policy.
    \item Assurance of operation -- ensures that the operational procedures stated in the guidelines are obeyed by the users. 
\end{itemize}
While the first three aspects are of technical character and can be determined by detailed inspection of the controls and policies, the last one is related to human factors.

\subsection{Rise of Adversarial Machine Learning}
Recently, there have been a rise of a new type of attacks called adversarial ML.
Adversarial attacks have different properties compared to traditional attack vectors.
In the ML area, adversarial attacks are a threat that is unique to ML models.
General frameworks described in Section~\ref{sec:infosec} were built before these attacks emerged and therefore were not designed to take them into consideration.
To be able to conduct risk analysis of adversarial ML, let us first classify the threats.
There are multiple classifications of attacks proposed up to date, either for general ML models~\cite{biggio2018wild} or on specialized classifiers, such as one proposed in~\cite{zhang2019generating} for text classifiers.

\begin{figure}
    \centering
    \includegraphics[width=0.6\textwidth]{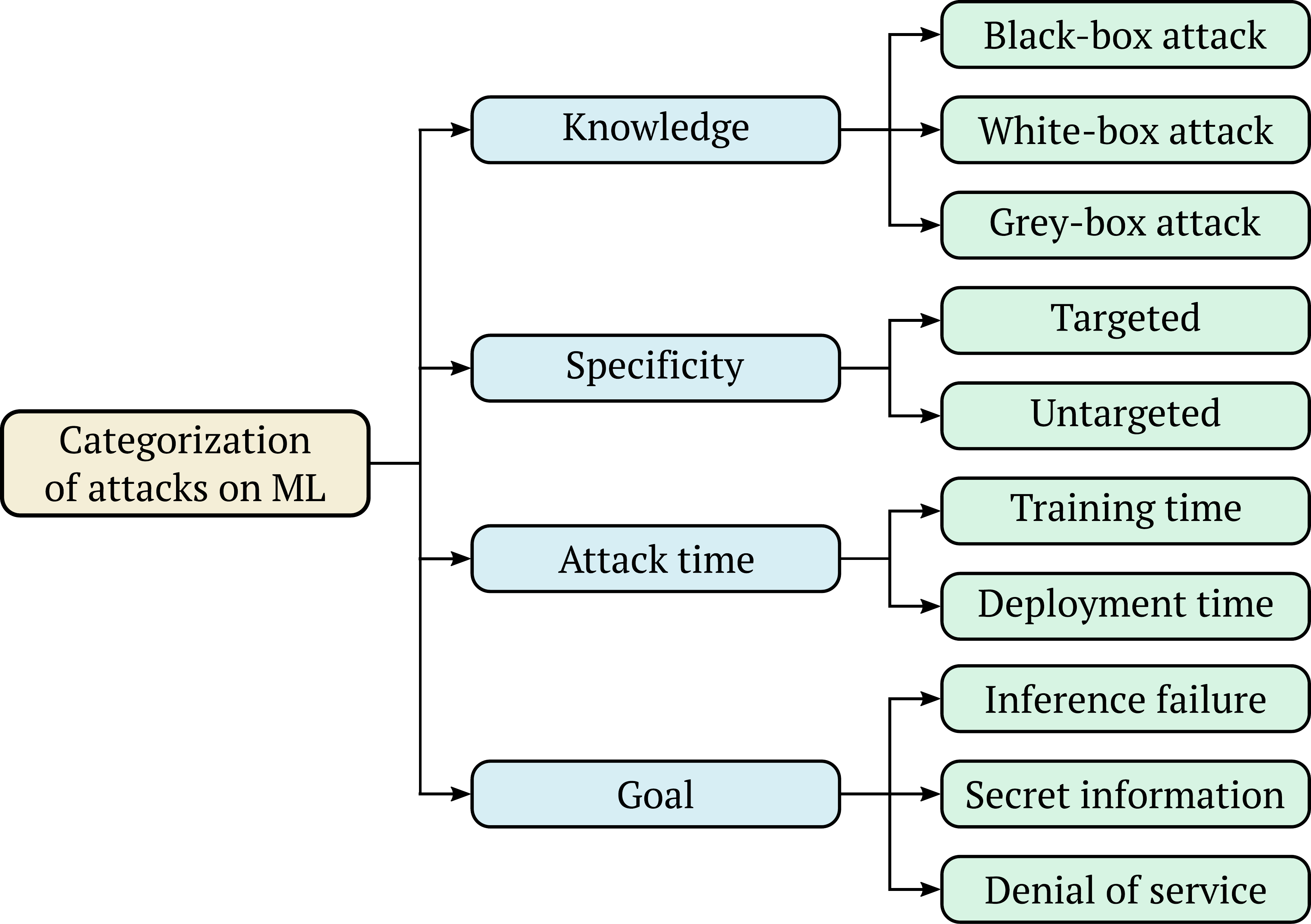}
    \caption{Classification of attacks on machine learning algorithms and models.}
    \label{fig:classification}
\end{figure}

In this paper, we propose to classify attacks into four major categories, based on:
\begin{itemize}
 \item \textbf{Attacker's knowledge} -- how much information about the model and data is available to the attacker.
 \item \textbf{Attack specificity} -- the amount that an attacker needs to manipulate the output of a model. For example, is it sufficient to cause a misclassification or does the attack need to target getting a particular classification result.
 \item \textbf{Attack time} -- the time and effort that the attacker uses in crafting their attack.
 \item \textbf{Attacker's goal} -- What is the attacker trying to achieve with a given attack. This may range from disrupting a business process; through to making particular gains from a business process to gaining a copy of a trained model.
\end{itemize}
Figure~\ref{fig:classification} depicts such classification.

When it comes to attack techniques, we categorize them according to attack surface into following:
\begin{itemize}
    \item \textbf{Data attacks} target the data that is used by the model. This can be either training (and test) data or data processed by the model during deployment. Following attack types are in this category:
    \begin{itemize}
        \item \textbf{Evasion attacks:} these attacks manipulate the input data to the model during the deployment~\cite{carlini2017towards}. For example, manipulation of malware code to have the corresponding sample misclassified as legitimate, or manipulation of images to mislead object recognition fall into this category. There are also attacks that focus on text based classifiers~\cite{liang2017deep}.
        \item \textbf{Poisoning attacks:} these attacks corrupt the classifier during the training time by adding a small number of \textit{poisoning samples} in the training set~\cite{biggio2012poisoning}. The misclassification during the deployment phase is then done by adding some feature in the input that is recognized by the poisoned classifier. An example of a targeted poisoning attack is a trojan~\cite{liu2017trojaning}.
        In this type of attack, the network is trained to misclassify into a specific class in case there is a trigger present in the input data.
        \item \textbf{Denial of service (DoS) attacks:} There are two sub-categories of DoS attacks: 1. Attacks intentionally providing data causing more resource consuming processing, thus causing the system run out of resources and ability to operate (e.g. sponge attack~\cite{shumailov2020sponge}); 2. Data availability attacks --  these attacks prevent the system from operation, effectively causing the denial of service.
        \item \textbf{Data corruption attacks:} these attacks cause the model to go into unpredictable state -- unable to produce the output. These are not novel attacks and work for any data-dependent system, however they might be more impactful on ML systems.
    \end{itemize}
    \item \textbf{Model attacks} target the model, focusing on affecting the confidentiality, integrity, and availability. Examples of these attacks are:
    \begin{itemize}
        \item \textbf{Model inversion attacks:} these attacks infer important features that were used as inputs to the model~\cite{fredrikson2015model}. For example, the attacker can recover a recognizable image of a person given only API access to a facial recognition system in the cloud and the name of the person whose face is recognized by it. Or an attacker may be able to infer confidential fields in tabular data when they have access to public fields possibly used in training.
        \item \textbf{Membership inference attacks:} these attacks aim at determining whether a certain record was used to train the target model~\cite{shokri2017membership}. Normally, training data is confidential information. Especially, in industrial settings, it is important to prevent the information leakage from a model that might be deployed outside of the organizational environment.
        \item \textbf{Model stealing attacks:} these attacks (sometimes called model extraction attacks) try to construct an identical or near-identical model to the original one~\cite{tramer2016stealing}. An adversary normally works in a black-box setting and crafts  specific queries to the model and observe the outputs. Based on this information, a surrogate model, that is is close agreement to the original, can be trained.
    \end{itemize}
    \item \textbf{Execution environment attacks} target the environment the model is executed in, either software or hardware. Examples of these attacks are:
    \begin{itemize}
        \item \textbf{Model corruption attacks:} these attacks tamper with the storage where the model is located. They can affect integrity and availability of the model.
    \end{itemize}
\end{itemize}

Additionally, we can look into attack methods.
As an ML model is a program running inside an environment that is deployed on a hardware system, the attacker has two ways to achieve the above-mentioned goals:
\begin{itemize}
    \item \textbf{Software attacks:} these attacks target the software itself. The majority of published works in adversarial domain are the software attacks. The assumption here is that the attacker makes either data changes, malicious queries, or information leaks on the software level.
    \item \textbf{Hardware attacks:} these attacks exploit the physical characteristics of the devices that are used to run the models. For example, a fault injection attack was used to achieve misclassification during deployment time~\cite{liu2017fault,breier2018practical}, and a side-channel attack was used to extract the model parameters~\cite{csi_nn}. Hardware attacks put stronger assumptions on the adversary as she has to be able to access the device physically.
    \item \textbf {Physical or Sensor Attacks}: Attacks can be made at the cyber-physical level such that the input into the digital environment running the model will lead to the wrong result. This could be as simple as changing a lens filter or placing a photo in front of a camera through to the placement of an adversarial patch next to an object being classified (leading to the wrong result)~\cite{brown2017adversarial}.
\end{itemize}

\subsection{Threat Modelling for ML}
There are several works aimed at identifying security risks in different parts of the ML lifecycle, for example Xiao et al.~\cite{xiao2018security} point at security flaws in popular deep learning frameworks such as TensorFlow, Torch, and Caffe.
Gu et al.~\cite{gu2017badnets} show security issues connected to outsourced training.
Bagdasaryan et al.~\cite{bagdasaryan2018backdoor} evaluate potential of poisoning attacks in federated learning.
Hayes and Ohrimenko~\cite{hayes2018contamination} discovered a problem in contamination of data in a multi-party models.

These, and many other attacks, show there are many attack vectors that need to be considered during the ML lifecycle, and therefore, a complex threat modelling is necessary when deploying ML models in an organizational environment.

The first ideas on threat modelling for ML were outlined by Barreno et al.~\cite{barreno2006can}. They defined the preliminary attack models that compromise integrity and availability of the models. At that time, ML models were mostly deployed in network security applications for anomaly detection. Therefore, the study mostly focused on models in intrusion detection systems. 

More recently, Kumar et al.~\cite{kumar2019failure} aimed at systematizing both intentional and unintentional failures in ML systems to help organizations tackle the potential issues connected to using ML models.
As they state, for example Microsoft modified their software development lifecycle process to include threat assessment of ML systems.

The question that remains is -- how do we take the existing threat modelling frameworks and incorporate the security issues that are unique to ML systems into them?
Current risk management standards are generic in terms of information systems and were mostly developed before organizations started using ML on a massive scale. 
Therefore, they do not capture the risks associated with data-driven computations with non-deterministic outcome. 
The goal of our framework is to bridge this gap and  provide an ML specific security framework for organizations dependent on ML methods. 

\subsection{Multiple-Criteria Decision Making}
Prioritizing different controls within the risk management framework is a multiple-criteria decision making (MCDM) problem~\cite{triantaphyllou2000multi}.
There are several techniques that can be used to for supporting the decisions~\cite{madm}:
\begin{itemize}
    \item \textit{Simple additive weighting (SAW)}, also known as weighted linear combination, is based on normalized weighted average. 
    Weights are assigned to each attribute and normalized to get the relative importance. These are then multiplied by the scaled value that is calculated by pairwise comparison between the alternatives~\cite{afshari2010simple}.
    \item \textit{Technique of order preference similarity to the ideal solution (TOPSIS)} first finds an ideal and anti-ideal solution. Then, it compares the distance of each solution to those. There are various ways that can be used in each step of TOPSIS, such as selection of the ideal and anti-ideal selection, distance metric, and normalization procedure~\cite{chen1992fuzzy}.
    \item \textit{Analytic hierarchy process (AHP)} is based on the relative measure theory, which does not consider the absolute performance of every solution, but rather relates performances between each two solutions. Similarly to SAW, pairwise comparisons are the basis of this method. The main usage is for cases where we are less interested in precise scores of alternatives and more in finding the best alternative among the available ones~\cite{saaty1988analytic}. 
    \item \textit{Preference Ranking Organization METHod for Enrichment of Evaluations (PROMETHEE)} ranks actions based on preference degrees, while the alternatives are assumed to be known according to the method~\cite{mareschal1984promethee}. It uses a pairwise comparison of actions based on selection criteria, computation of unicriterion flows, and aggregation of those into global flows. It was applied to risk assessment before~\cite{zhang2009comparative}.
\end{itemize}
In our work we utilize a variation of SAW to derive the weights for attributes within the risk management framework.
We omit the pairwise comparison to keep the method easy to use for the evaluator -- otherwise, the number of inputs might be too high for a practical usage.
It is possible to adjust the MCDM technique or select an alternative one based on preference or use case and utilize the other parts of the framework.

\section{Risk Management Framework for Machine Learning Based Solutions}
\begin{figure}
    \centering
    \includegraphics[width=0.45\textwidth]{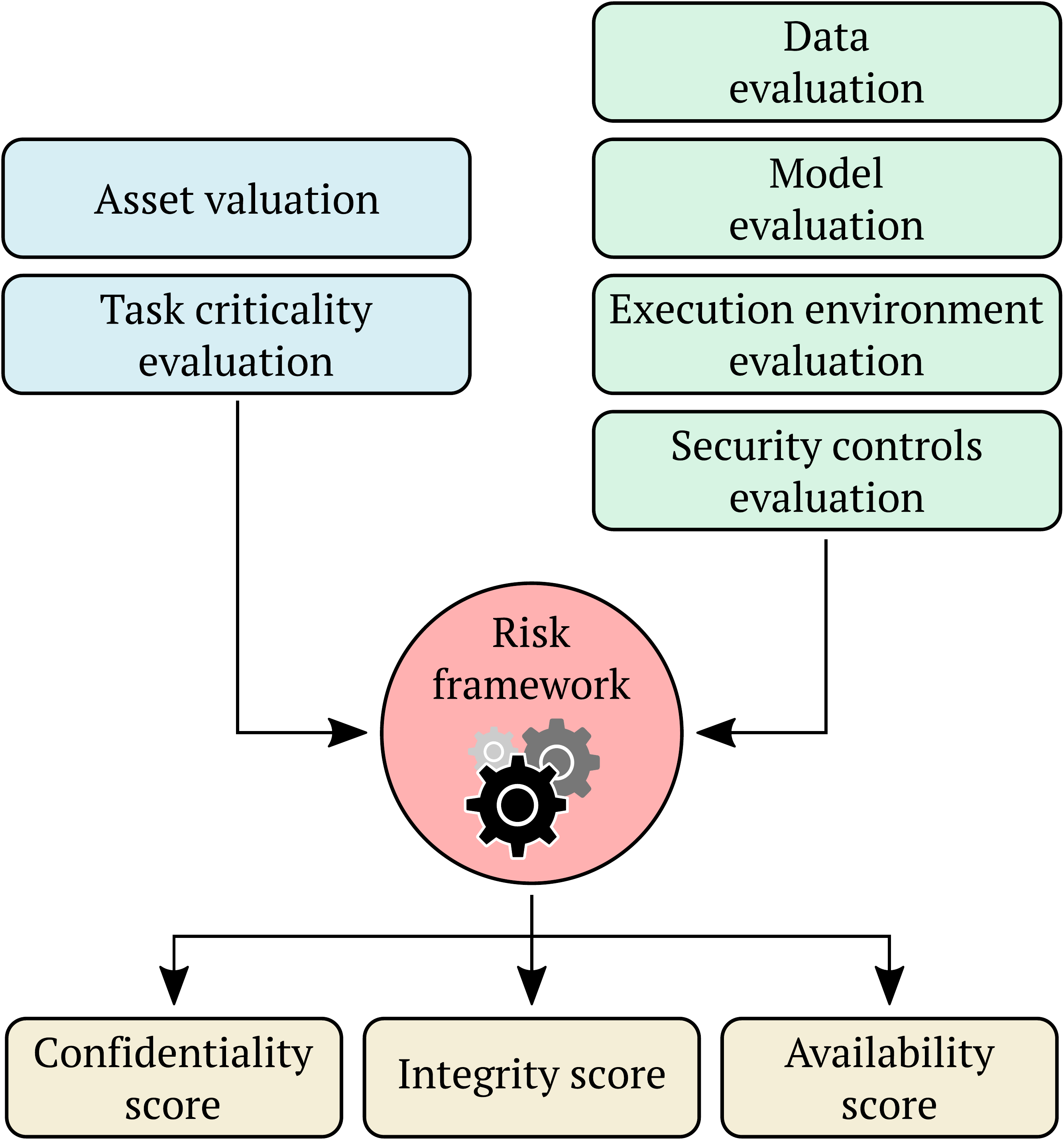}
    \caption{Risk management framework overview diagram.}
    \label{fig:framework}
\end{figure}

The risk management framework for ML security in this section follows a similar approach to the traditional risk frameworks stated in Section~\ref{sec:infosec}.
A high-level overview of the framework is depicted in Figure~\ref{fig:framework}.
The left side inputs (blue color) provide information on how critical the task is and what is the value of assets that are at risk.
The right side inputs (green color) are results of an evaluation, as guided by the framework, that was carried on the data, model, and security controls that protect the entire solution lifecycle.
The framework takes these inputs and provides an output in a form of confidentiality, integrity, and availability scores.
Depending on how the organization's security policy point seeks to balance availability, confidentiality and integrity, risk mitigation can be carried out to minimize the risks in the appropriate areas. 

\paragraph{Assets and tasks.} Normally, ML models are used to perform a certain task in the organizational processes and in doing so they affect certain assets in the organizational environment.
These assets and tasks might be at risk due to the usage of ML models, depending on the level of implemented security measures.
It is therefore important to identify the tasks and the assets to understand how these can be affected -- there might be tasks leading to financial cost, poor quality control, disclosure of confidential information, corruption, limited availability or higher cost associated with the potential threat.

\paragraph{Evaluation factors (EF).} To determine whether the protection of assets and tasks is adequate to organization's mission, we look at evaluation factors.
These factors are the controls implemented to increase the security state within the organization and while some of them are relatively general in their nature (security controls EFs), the others are specific to machine learning area (data/model EFs).
When looking at EFs, we are interested in getting knowledge of:
\begin{itemize}
    \item \textbf{Implementation quality} -- this is done by evaluating whether an EF is in place and if proper procedures are followed to make an EF effective.
    \item \textbf{Coverage} -- this is to determine the importance of an EF. Coverage simply means how many assets and tasks are influenced by an EF. Then, based on the value and criticality of these assets and tasks, it is possible to identify critical EFs, and also weak areas that require attention.
\end{itemize}
Each EF can contribute to overall ML lifecycle security  either in a \textit{proactive} or \textit{reactive} way, or both.
In the proactive way, an EF is actively preparing the system for a potential failure, for example by performing automatic backups.
In the reactive way, an EF acts in case there is a security incident, for example by doing a forensic investigation in the network.
As the factors are divided into groups according to the risk management framework design, it is possible to determine relative importance of each factor's contribution to C/I/A score in the proactive and reactive domains.
By doing that, it is easier for decision makers to prioritize controls that contribute significantly to security attributes that are the most crucial for the organization.
During the evaluation, discrete values are assigned to each evaluation factor based on their implementation quality.

\subsection{Evaluation Method}
\begin{figure}
    \centering
    \includegraphics[width=0.56\textwidth]{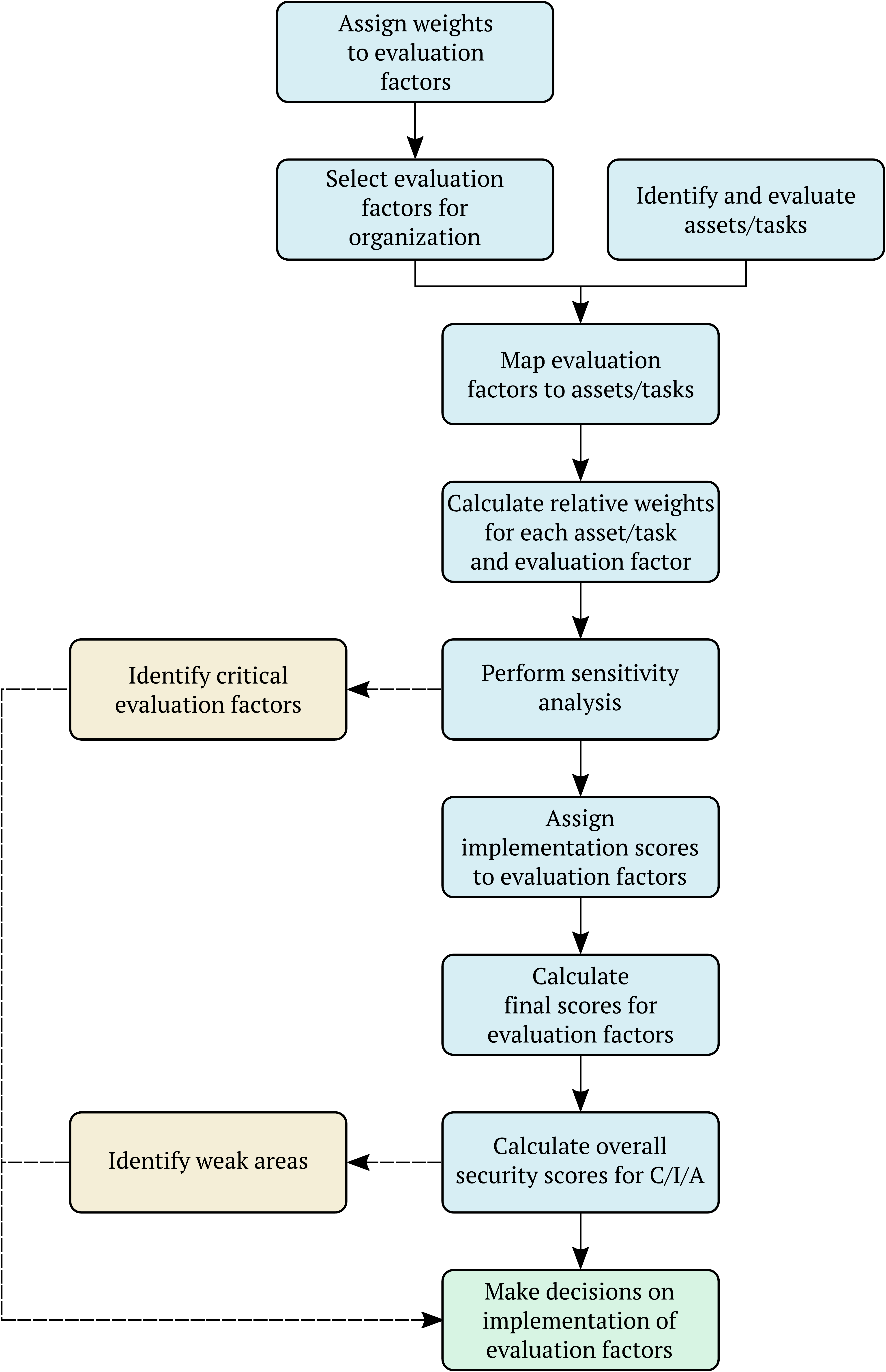}
    \caption{Evaluation process with the risk management framework.}
    \label{fig:framework_flow}
\end{figure}

The entire process can be described by following steps (for better clarity, the flow graph of the process is depicted in Figure~\ref{fig:framework_flow}):
\begin{enumerate}
    \item Each evaluation factor is assigned a base weight on how well it contributes to each security attribute (C/I/A) in proactive and reactive way on a discrete scale from 0 (no contribution) to 5 (maximum contribution).
    \item Tailoring of evaluation factors to the organizational environment -- in case some EFs are not applicable (such as \textit{Labeling quality} in case of unsupervised tasks), they can be removed from the evaluation. This step has to be well-documented and valid reasons need to be given.
    \item Assets and tasks that are influenced by the usage of ML models are identified and evaluated. Each of them is given a discrete value between 1 and 100 to express their importance to the organization. The numbers are later normalized to get the relative importance of assets/tasks.
    \item Mapping of evaluation factors to assets and tasks is done to calculate the coverage. Each EF can contribute to each asset/tasks on a discrete scale between 0 to 5. 
    \item Using the formula in Equation~\ref{eq:weight}, relative weight for contribution of each EF to each asset/task is calculated. This also reflects the categorization of EFs in different evaluation groups (data/model/security controls).
    \item Sensitivity analysis -- implementation scores of relevant EFs are varied to get the overall picture of which EFs are contributing the most to protecting the organization's assets and tasks. Based on this, the organization can determine which EFs have to be evaluated with high degree of precision, as even slight change could strongly affect the final scores in terms of security attributes.
    \item Implementation score is assigned to each evaluation factor based on the values in the Tables~\ref{app:ef_data_eval},\ref{app:ef_model_eval},\ref{app:ef_exec_eval} and \ref{app:ef_sec_eval}. It is important to note that any value in interval $[0,1]$ can be assigned, the three values in the tables serve only as a guideline. 
    \item Implementation score of each evaluation factor is multiplied with its relative weight related to an asset to get the final score.
    \item Final scores of all the evaluation factors are summed up within each security attribute (C/I/A) and category (proactive/reactive) to get the overall security score.
    \item The assessor then looks at whether the risk is sufficiently mitigated for the business given the results of the calculation (or they could identify too little risk and operating the controls would be too expensive). The assessor can then use the tool to explore what are the most cost effective ways to improve the risk situation. 
\end{enumerate}
It has to be noted that the first step is ``offline'' step -- it is not a part of the evaluation as the base weights calculated for each evaluation factor are fixed and should remain the same across different evaluations.

Overview of evaluation factors split into categories is depicted in Figure~\ref{fig:efs}. Full description of each evaluation factor is stated in Appendices~\ref{app:sec_ef_data}-\ref{app:sec_ef_controls}.

\begin{figure*}
    \centering
    \includegraphics[width=0.98\textwidth]{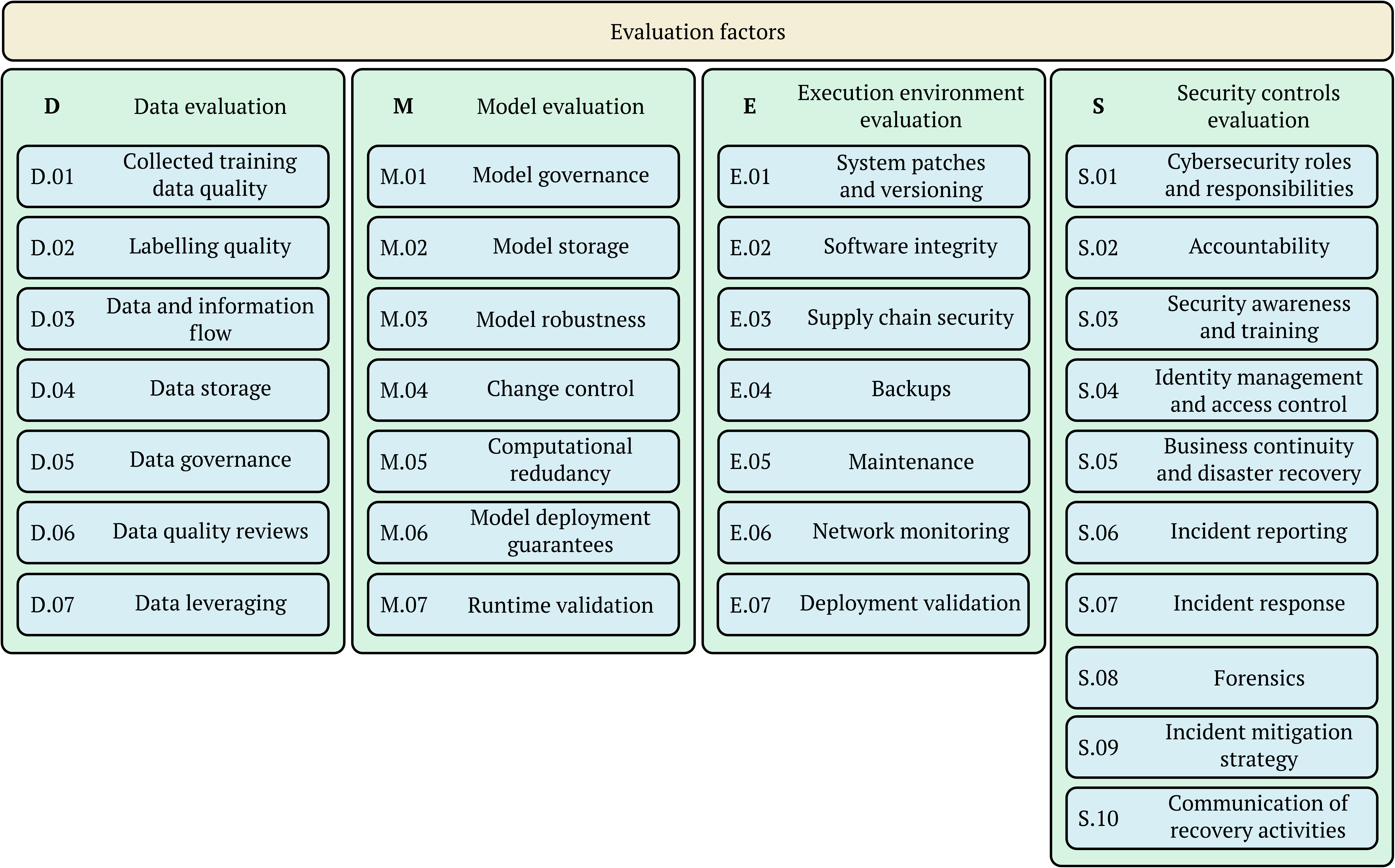}
    \caption{Evaluation factors split into categories -- data evaluation, model evaluation, execution environment evaluation, and security controls evaluation. Full description of each evaluation factor is stated in Appendices~\ref{app:sec_ef_data}-\ref{app:sec_ef_controls}.}
    \label{fig:efs}
\end{figure*}

\subsection{Evaluation Formulas}
In this part, we provide the details of the calculations for different components of the framework.

\paragraph{Relative weights} Let us denote the base weight for $EF_i$ as $BW_{T,D}^{EF_i} \in [0,5]$, where $T \in \{C,I,A\}$ is the security attribute (confidentiality, integrity, availability) to which the weight corresponds, and $D \in \{P,R\}$ is the domain that can be either proactive or reactive.
Let us denote the mapping of an evaluation factor $EF_i$ to an asset $A_j$ by $M(A_j, EF_i) \in [0,5]$.
Now, we can derive the formula for the relative weight contributions as:
\begin{equation}\label{eq:weight}
\resizebox{0.7\textwidth}{!}{$
W^{A_j,EF_i}_{T,D} = 
\begin{cases}
      \displaystyle\frac{BW_{T,D}^{EF_i} \times M(A_j,EF_i)}{\displaystyle\sum_{k=1}^n BW_{T,D}^{EF_k} \times M(A_j,EF_k)}, & \text{if}\ \sum_{k=1}^n BW_{T,D}^{EF_k} \times M(A_j,EF_k) \neq 0 \\
      0, & \text{otherwise}
    \end{cases}$}
\end{equation}
where $n$ is the number of mappings between evaluation factors for given security attribute and asset $A_j$ in domain $D$.

\paragraph{Protection scores} Now, after we have all the relative weights $W$ and all the implementation scores $S$ for evaluation factors, we can proceed with the final evaluation.
Protection score $P^{A_j}_{T,D}$ for asset $A_j$ can be calculated as follows:
\begin{equation}\label{eq:protection}
P^{A_j}_{T,D} = \sum_{i=1}^{n} W^{A_j,EF_i}_{T,D} \times S_{EF_i}
\end{equation}
\textbf{Final score} for security attribute $T$ in the domain $D$ are then simply calculated as a sum of the protection scores for all the assets:
\begin{equation}\label{eq:final}
FS_{T,D} = \frac{\sum_{j=1}^{m} P^{A_j}_{T,D}}{m}
\end{equation}
where $m$ is the number of assets.

\paragraph{Security coverage} Now, after we have the security scores, it is also important to know how does it translate in terms of asset values.
Let us denote the value of asset $A_j$ as $V_{A_j}$.
Security coverage $C^{A_j}_{T,D}$ of an asset $A_j$ for security attribute $T$ in domain $D$ can be then calculated as:
\begin{equation}\label{eq:coverage}
C^{A_j}_{T,D} = P^{A_j}_{T,D} \times V_{A_j}
\end{equation}
Total coverage of all the organization's assets, taking account of their value, can be calculated as:
\begin{equation}\label{eq:total_coverage}
TC_{T,D} = \sum_{j=1}^{m} C^{A_j}_{T,D}
\end{equation}
Total coverage will indicate how well are the organization's assets protected on a scale of  $[0,1]$.

\paragraph{Thresholds.} It is possible to set thresholds for either individual evaluation factors, assets, or evaluation groups. 
This is important in case there is a certain baseline that requires balanced security in all the areas.

\subsection{Cost of evaluation factors}
Another perspective to look at is the cost (initial and operational) of the controls vs. their contribution to security attributes and to relate this to  asset value and task criticality.
If there are new controls that require implementation, it is important to know the RoI and to assess to relative cost effectiveness of different control choices.

\begin{figure}
    \centering
    \includegraphics[width=0.4\textwidth]{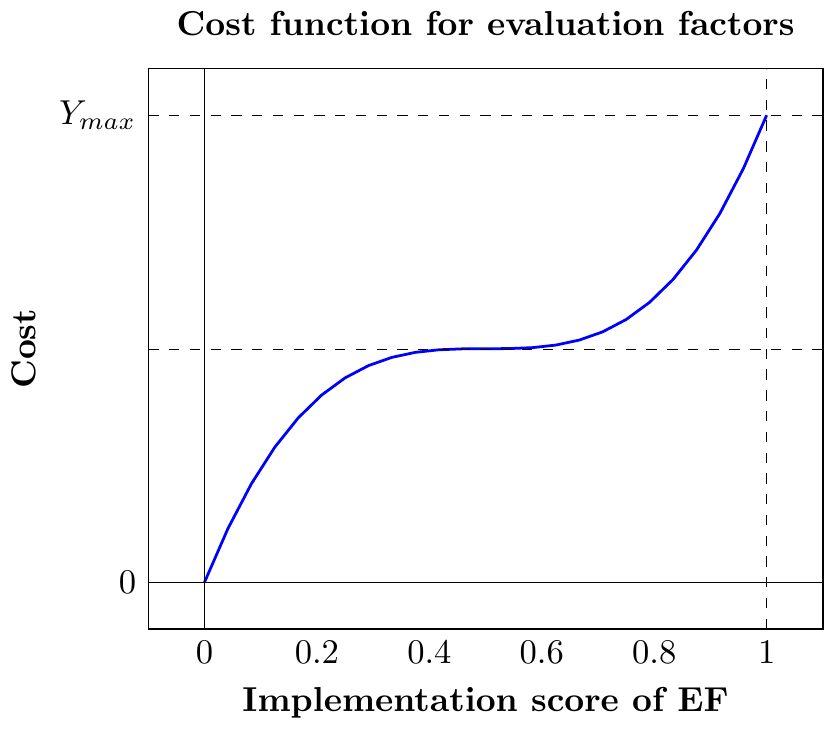}
    \caption{Cost function for evaluation factors. Non-linear shape captures the increasing cost at the beginning and at the end.}
    \label{fig:cost}
\end{figure}

When it comes to implementing evaluation factors, the cost corresponding to the evaluation score is not linear.
Usually, there is a rapid growth in cost at the beginning when an EF is selected to be implemented, and then the cost saturates around the middle levels where small improvement does not increase it too much.
However, when there is a requirement to have a very high implementation score, the cost starts growing rapidly as well.
This phenomenon is depicted in Figure~\ref{fig:cost} which plots the cost function designed for our evaluation method.
Below, we describe this function more formally.

\paragraph{Current cost} Let us denote the maximum cost for $EF_i$ as $Y^{EF_i}_{max}$ (this is the cost when $S_{EF_i} = 1$). 
Now we can derive the formula for calculating the partial cost $Y^{EF_i}$ of $EF_i$ that depends on its implementation score $S_{EF_i}$:
\begin{equation}\label{eq:cost}
Y^{EF_i} = \big((2\times S_{EF_i}-1)^3+1\big) \times \frac{Y^{EF_i}_{max}}{2}
\end{equation}

\paragraph{Cost efficiency} The question that comes when deciding on implementation score of an EF, is how to find the right balance between security investments and the security state of the organization.
To do this, we can minimize the ratio between the current cost of all the evaluation factors while varying their implementation score, and the total coverage of organization's assets.
In this step, one also has to decide what components of the total coverage should be included -- proactive/reactive domain, confidentiality/integrity/availability security attributes.
For example, for balanced security, a sum of all the components (6 in total) can be chosen.
The formula to find the optimal efficiency $E$ is as follows:
\begin{equation}\label{eq:efficiency}
E = min_{S_{EF_i}}\bigg(\frac{\sum^{n}_{i=1}Y^{EF_i}}{TC_{sel}}\bigg)
\end{equation}
where $TC_{sel}$ is the sum of total coverages for selected components.
In other words, we look at all the possible implementation scores of EFs, calculate their current cost, sum these costs, and divide by the total coverage that corresponds to the evaluated implementation scores.
If the weights are properly assigned, this gives us a good approximation on where to invest in terms of evaluation factors.
It is important to note that because of high number of EFs in the model, it might be computationally infeasible to enumerate all the possible values for their implementation scores.
Therefore, we recommend usage of some optimization methods~\cite{cavazzuti2012optimization}.

\paragraph{Government requirements} Apart from the cost directly associated with the breach, there are also costs associated with the legal requirements.
For example, systems that process personal information, or even sensitive personal information (such as health records), need to have protection measures in place in most countries according to the law.
If an organization is being audited and it fails to meet the legal requirements, the associated fines may be significant.
For example, violators of EU General Data Protection Regulation (GDPR) may face fines up to 20 million EUR, or up to 4\% of the annual worldwide turnover from the preceding financial year~\cite{gdpr}.
Although we do not include these costs in our model, it is necessary to consider them when designing on the security investments.

\section{Example}
Let us consider a simple organizational example with three evaluation factors and four assets to provide an overview on how our method works.
For the sake of simplicity, we will not consider tasks in this example, as the evaluation would work in the same way as for assets.

The three evaluation factors ($EF_{1-3}$) with their corresponding base weights are stated in Table~\ref{tab:example}.
The cost of these factors are $EF_1 = 15.000, EF_2 = 30.000, EF_3 = 12.000$ units.
Let us first look into security properties of these factors to see how each of them contributes to security attributes:
\begin{itemize}
    \item $EF_1$: contributes to availability only, mostly in a proactive way. It can be, for example, a load balancing component or something similar. 
    \item $EF_2$: in a proactive domain, it contributes to all three attributes, but relatively weaker than $EF_3$ when looking at confidentiality and integrity, and also weaker than $EF_1$ when considering the availability. In a reactive domain, it strongly contributes to confidentiality and integrity. This can be for example some complex component such as data storage which provides redundancy, strong authentication, and accountability.
    \item $EF_3$: contributes to confidentiality and integrity attributes, where it is focused mostly on proactive domain. This could be for example robust model training.
\end{itemize}

\begin{table}
\caption{Evaluation factors' base weights used in example.}
    \label{tab:example}
    \centering
    \begin{tabular}{|c|ccc|ccc|}\cline{2-7}
    \multicolumn{1}{c|}{} & \multicolumn{6}{c|}{Base weights}\\ \cline{2-7}
        \multicolumn{1}{c|}{} & \multicolumn{3}{c}{Proactive} & \multicolumn{3}{|c|}{Reactive} \\\hline
        Evaluation factor & C & I & A & C & I & A \\ \hline\hline
        $EF_1$ & 0 & 0 & 4 & 0 & 0 & 2\\
        $EF_2$ & 2 & 3 & 1 & 4 & 3 & 0\\
        $EF_3$ & 4 & 4 & 0 & 2 & 2 & 0\\\hline
    \end{tabular}
\end{table}

Now, let us have a look at assets and their value to the organization. 
Table~\ref{tab:example_assets} shows the base value and normalized value of the assets -- aiding the recognition of their relative importance to the organization.

\begin{table}
\caption{Assets used in the example and their value.}
    \label{tab:example_assets}
    \centering
    \begin{tabular}{|c|c|c|c|c|}\hline
        Asset &  $A_1$ & $A_2$ & $A_3$ & $A_4$\\ \hline\hline
        Value &   45 & 10 & 35 & 75 \\
        Normalized value & 0.28 & 0.06 & 0.21 & 0.46 \\\hline
    \end{tabular}
\end{table}

\begin{table}
\caption{Mapping of evaluation factors to assets.}
    \label{tab:example_assets_efs}
    \centering
    \begin{tabular}{|c|c|c|c|c|}\hline
        Asset &  $A_1$ & $A_2$ & $A_3$ & $A_4$\\ \hline\hline
        $EF_1$ &   1 & 0 & 2 & 4 \\
        $EF_2$ &   2 & 2 & 1 & 1 \\
        $EF_3$ &   5 & 0 & 1 & 0 \\\hline
    \end{tabular}
    
\end{table}

\begin{table}
\caption{Calculation of the relative weights of contribution to C/I/A by evaluation factors.}
    \label{tab:example_final}
    \centering
    \begin{tabular}{|c|ccc|ccc|ccc|ccc|}\cline{2-13}
        \multicolumn{1}{c|}{} &  \multicolumn{12}{c|}{Assets} \\ \cline{2-13}
        \multicolumn{1}{c|}{} &  \multicolumn{6}{c|}{$A_1$} & \multicolumn{6}{c|}{$A_2$} \\ \cline{2-13}
        \multicolumn{1}{c|}{} &  \multicolumn{3}{c|}{Proactive}  & \multicolumn{3}{c|}{Reactive} & \multicolumn{3}{c|}{Proactive}  & \multicolumn{3}{c|}{Reactive} \\ \hline
        EFs & C & I & A & C & I & A & C & I & A & C & I & A  \\ \hline\hline
        $EF_1$ &   0    & 0    & 0.67 & 0    & 0    & 1.00 & 0   & 0   & 0   & 0   & 0   & 0 \\
        $EF_2$ &   0.17 & 0.23 & 0.33 & 0.44 & 0.38 & 0   & 1.00 & 1.00 & 1.00 & 1.00 & 1.00 & 1.00 \\
        $EF_3$ &   0.83 & 0.77 & 0    & 0.56 & 0.62 & 0   & 0   & 0   & 0   & 0   & 0   & 0\\\hline
    \end{tabular}
    \begin{tabular}{|c|ccc|ccc|ccc|ccc|}\cline{2-13}
        \multicolumn{1}{c|}{} &  \multicolumn{12}{c|}{Assets} \\ \cline{2-13}
        \multicolumn{1}{c|}{} &  \multicolumn{6}{c|}{$A_3$} & \multicolumn{6}{c|}{$A_4$} \\ \cline{2-13}
        \multicolumn{1}{c|}{} & \multicolumn{3}{c|}{Proactive} & \multicolumn{3}{c|}{Reactive}  & \multicolumn{3}{c|}{Proactive}  & \multicolumn{3}{c|}{Reactive}  \\ \hline
        EFs & C & I & A & C & I & A & C & I & A & C & I & A   \\ \hline\hline
        $EF_1$ & 0    & 0      & 0.89 & 0    & 0    & 1.00 & 0   & 0   & 0.94 & 0 & 0 & 1.00\\
        $EF_2$ & 0.33 & 0.43 & 0.11 & 0.57 & 0.60 & 0   & 1.00 & 1.00 & 0.06 & 1.00 & 1.00 & 0\\
        $EF_3$ & 0.67 & 0.57 & 0    & 0.43 & 0.40 & 0   & 0   & 0   & 0    & 0 & 0 & 0\\\hline
    \end{tabular}
\end{table}

\begin{figure}
    \centering
    \begin{tabular}{cc}
        \includegraphics[width=0.45\textwidth]{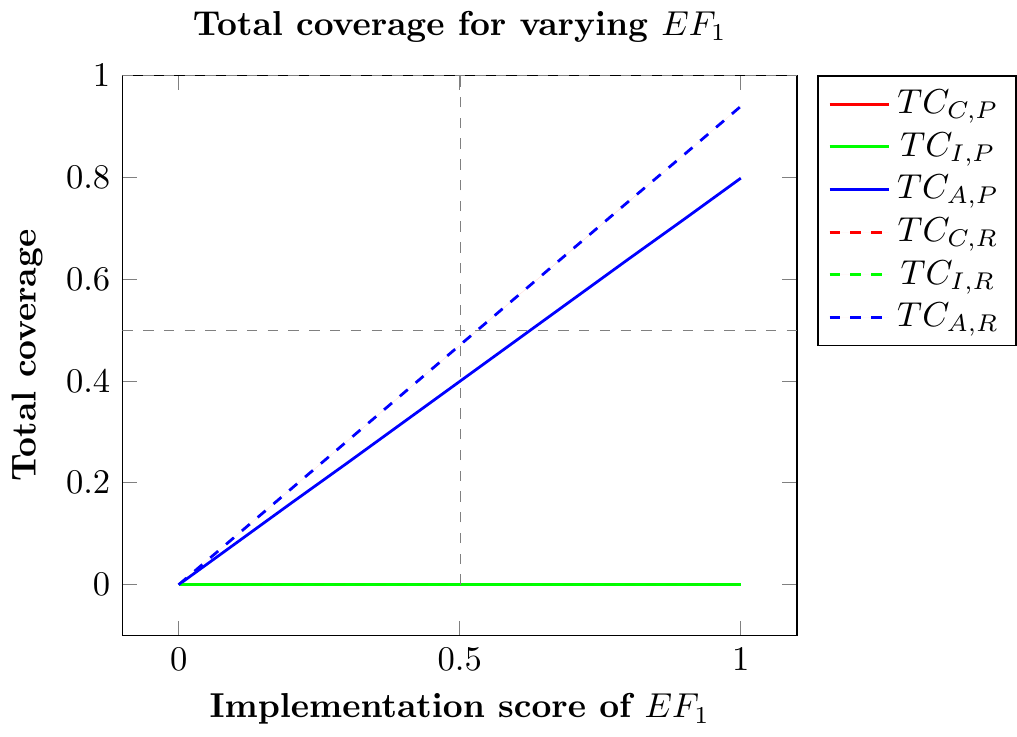} & \includegraphics[width=0.45\textwidth]{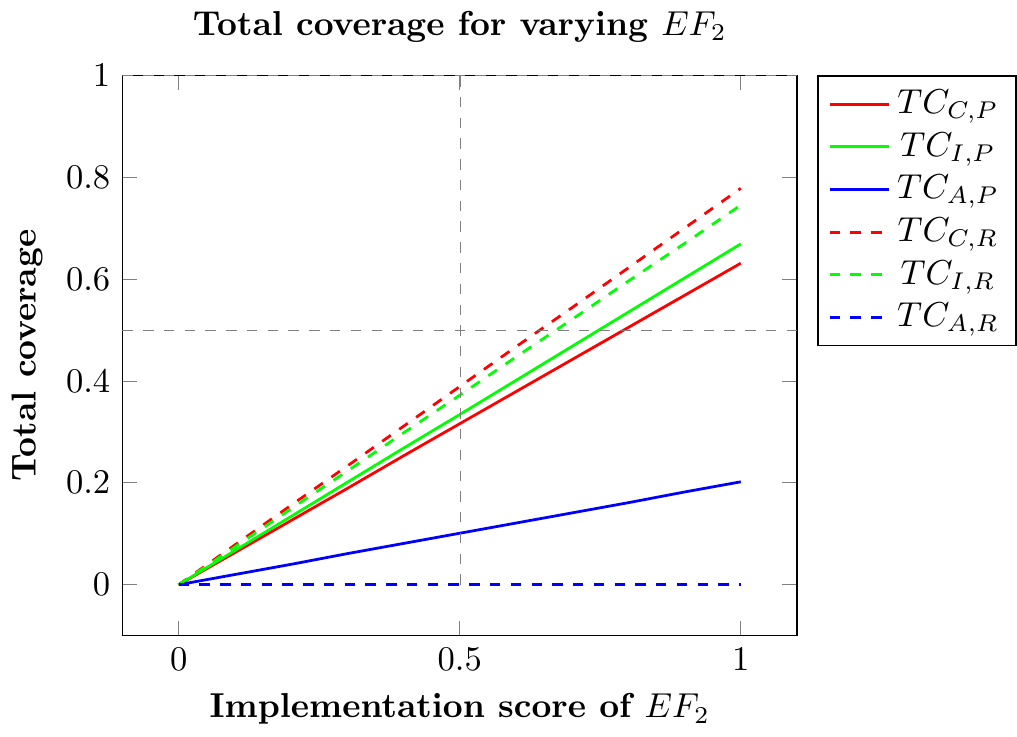}\\
        (a) & (b) \\
        \multicolumn{2}{c}{\includegraphics[width=0.45\textwidth]{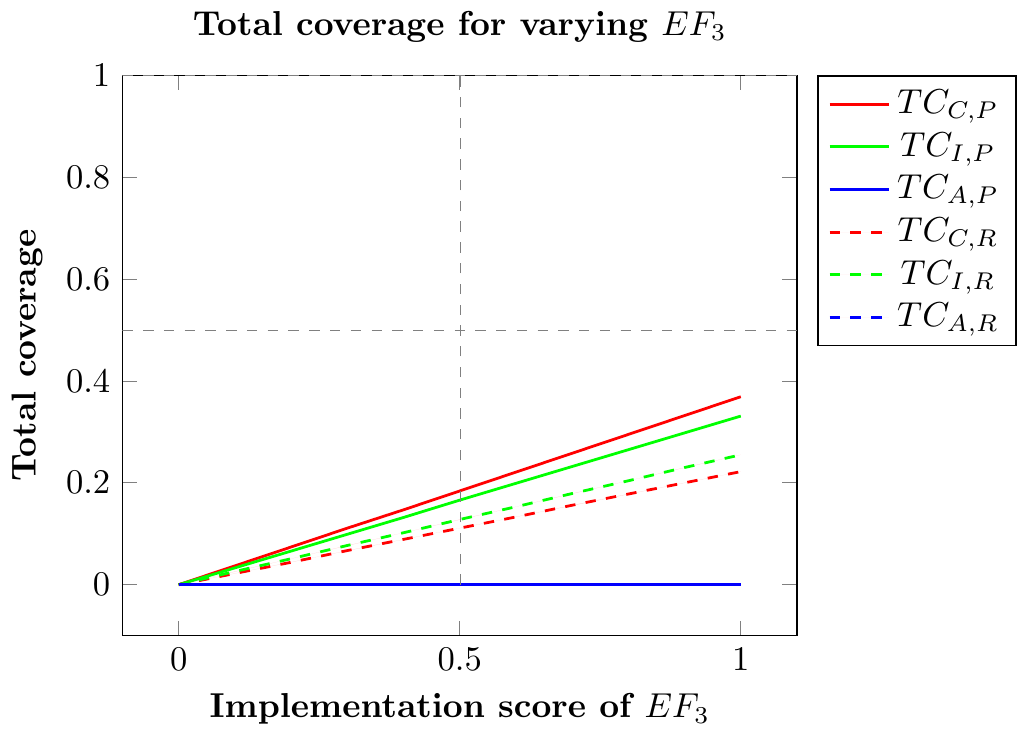}}\\
         \multicolumn{2}{c}{(c)}
    \end{tabular}
    \caption{Sensitivity analysis for the example showing total coverage for varying each EF separately: (a) $EF_1$, (b) $EF_2$, and (c) $EF_3$.}
    \label{fig:EF_contribution}
\end{figure}

After knowing the characteristics of the EFs and the value of assets, we can map the EFs to assets.
This part is depicted in Figure~\ref{tab:example_assets_efs}.
Mapping and base weights give us the values of relative weights according to Equation~\ref{eq:weight}.
These values are stated in Figure~\ref{tab:example_final}.
Now, we can perform a sensitivity analysis step for each EF and measure how it contributes to  overall security.
We plot the sensitivity analysis in Figure~\ref{fig:EF_contribution} where we vary the implementation score of each EF separately and measure the total coverage.
These plots show us the trade-offs that can be done when implementing the controls to protect the organization.
For example, if we are focused on all the three attributes, it makes sense to fully implement $EF_2$ and then partially implement either $EF_1$ or $EF_3$ based on whether confidentiality+integrity is more important, or availability.
In case we  care more about availability, $EF_1$ is the best choice, and the resources can be saved on not implementing the other two.
Another observation, under an assumption of limited security resources, is that, in the case where integrity is the most crucial attribute (which it often is, for providing safety features to ML models) while the other two attributes are moderately important, it might be useful to partially implement $EF_2$ and $EF_3$ rather than only fully implementing $EF_3$.
In such case, the integrity score would be modest, but the $EF_2$ would also contribute to the other two attributes. 

\begin{figure}
    \centering
    \begin{tabular}{cc}
        \includegraphics[width=0.45\textwidth]{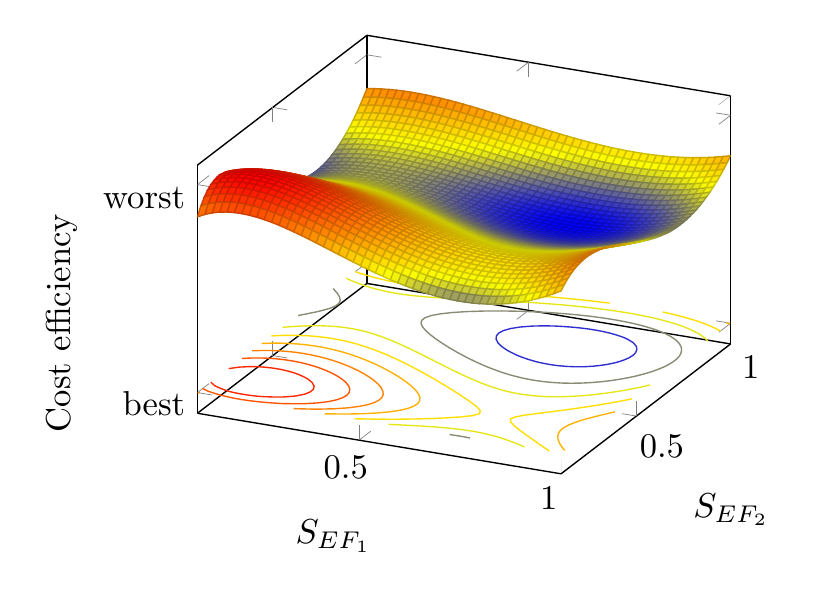} & \includegraphics[width=0.45\textwidth]{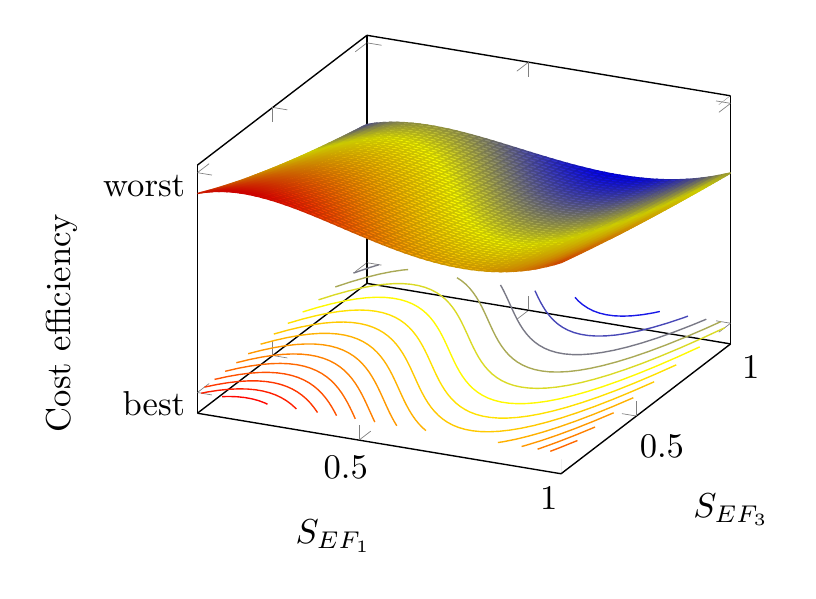} \\
        (a) & (b)\\
        \multicolumn{2}{c}{\includegraphics[width=0.45\textwidth]{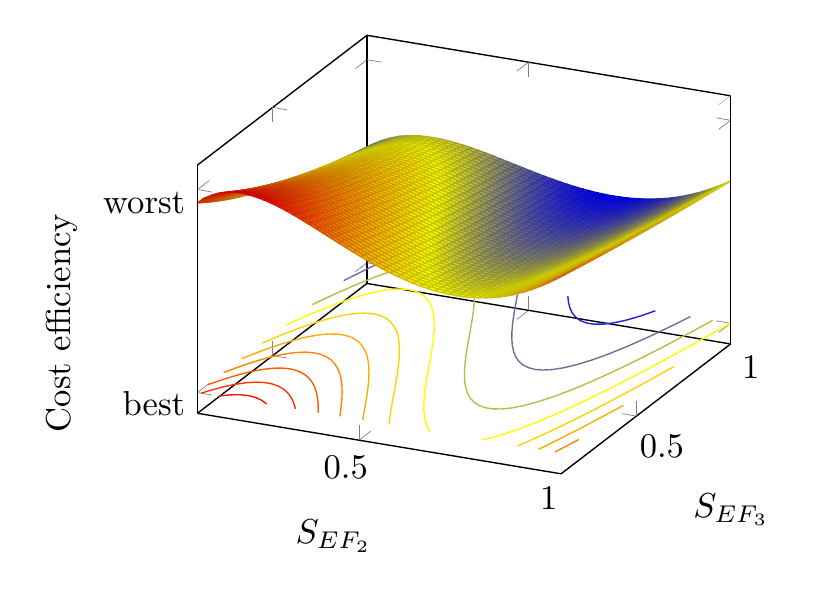}}\\
         \multicolumn{2}{c}{(c)}
    \end{tabular}
    \caption{Cost efficiency for the three EFs. Plot (a) varies $S_{EF_1}$ and $S_{EF_2}$, (b) varies $S_{EF_1}$ and $S_{EF_3}$, (c) varies $S_{EF_2}$ and $S_{EF_3}$. The remaining EF implementation score is always set to 0.7.}
    \label{fig:EF_efficiency}
\end{figure}

\paragraph{Cost modelling of EFs} An important part on deciding on how well to implement the selected EFs is to check the cost efficiency -- investment w.r.t. total coverage.
This can be done by varying the implementation score of each EF separately and calculating the cost efficiency according to Equation~\ref{eq:efficiency}.
As we have three EFs in this example, it is possible to enumerate all the possible values of implementation scores.
Let us assume we want to implement each EF at least at a basic level, i.e. the implementation score is at least $0.1$.
After the enumeration, we can see that the most cost efficient solution is to have the implementation scores as follows: $S_{EF_1} = 0.8, S_{EF_2} = 0.7, S_{EF_3} = 0.7$. 
To visualize the cost efficiency, we show 3D plots in Figure~\ref{fig:EF_efficiency}.
Each plot shows cost efficiency with one EF fixed to implementation score of 0.7, while varying the other two EFs.

\section{Conclusion}
In this paper, we have proposed a risk management framework tailored for organizations that are reliant on using machine learning models in their processes.
Our framework is based on identifying critical assets and tasks and mapping evaluation factors to them.
Evaluation factors show the implementation scores of the security measures that are in place and protect the confidentiality, integrity, and availability of assets and tasks, either in reactive or proactive way, or both.
We believe our framework can be helpful in determining the weak areas of security and in guiding decision makers in implementing proper security measures.

\section*{Acknowledgement}

\noindent
This work has been supported in parts by the ``University SAL Labs'' initiative of Silicon Austria Labs (SAL) and its Austrian partner universities for applied fundamental research for electronic based systems.

\bibliographystyle{plain}
\bibliography{references}

\appendix

\section{Data Evaluation Factors}
\label{app:sec_ef_data}

\begin{center}\footnotesize
\tablecaption{Data evaluation factors defined in the framework.} 
 \label{app:ef_data_eval}
\begin{supertabular}{|p{3cm}|p{9.5cm}|}\hline 
    \multicolumn{2}{|c|}{\textbf{D.01 Collected training data quality}}\\\hline
        No guarantees & data provenance is not clear, integrity cannot be checked\\\hline
        Partial guarantees & data provenance is known, integrity can be checked to some extent\\\hline
        High guarantees & data provenance is known, all the data is inventoried, integrity is guaranteed\\\hline
    \hline
    \multicolumn{2}{|c|}{\textbf{D.02 Labeling quality}}\\\hline
        No guarantees & labeling of the data was done by a third-party and was not checked for correctness\\\hline
        Partial guarantees & labeling was done by a trustworthy third-party, but was not checked for correctness\\\hline
        High guarantees & all the data labeling was checked for correctness\\\hline
        \hline
    \multicolumn{2}{|c|}{\textbf{D.03 Data and information flow}}\\\hline
        Not secure & data flow to the model is not secured by information security controls -- confidentiality, integrity, and availability can be compromised\\\hline
        Secure & confidentiality and integrity is protected by security controls, but availability can be affected\\\hline
        Secure and guaranteed & like Secure, but with added measures to protect availability to support business continuity\\\hline
        \hline
    \multicolumn{2}{|c|}{\textbf{D.04 Data storage}}\\\hline
        Not secure & data storage is not secured by information security controls -- confidentiality, integrity, and availability can be compromised\\\hline
        Secure & confidentiality and integrity is protected by security controls, but availability can be affected\\\hline
        Secure and guaranteed & like Secure, but with added measures to protect availability to support business continuity\\\hline
        \hline
    \multicolumn{2}{|c|}{\textbf{D.05 Data governance}}\\\hline
        Undefined & data governance is not clear -- accountability is not implemented\\\hline
        Semi-defined & data governance is defined but the responsibilities are not clear\\\hline
        Defined & responsibilities for handling and securing the data are clearly defined and assigned to trained personnel\\\hline
        \hline
    \multicolumn{2}{|c|}{\textbf{D.06 Data quality reviews}}\\\hline
        Undefined & there are no reviews related to data quality\\\hline
        Ad-hoc & data quality is reviewed in case a problem is found\\\hline
        Regular & policies to guarantee regular data quality reviews are deployed\\\hline
        \hline
    \multicolumn{2}{|c|}{\textbf{D.07 Data leveraging}}\\\hline
        Undefined & there is no process in place to ensure proper usage of data from public domain or other sources\\\hline
        Semi-defined & there are guidelines in place to handle usage of data from public domain and other sources but their application is not strictly enforced \\\hline
        Defined & there is a formal process in place to handle usage of data from public domain and other sources\\\hline
\end{supertabular}
\end{center}

\section{Model Evaluation Factors}
\label{app:sec_ef_model}

\begin{center}\footnotesize
\tablecaption{Model evaluation factors defined in the framework.} 
 \label{app:ef_model_eval}
\begin{supertabular}{|p{3cm}|p{9.5cm}|}\hline
    \multicolumn{2}{|c|}{\textbf{M.01 Model governance}}\\\hline
        Undefined & model governance is not clear -- accountability is not implemented\\\hline
        Semi-defined & governance ownership is defined but the responsibilities are not clear\\\hline
        Defined & responsibilities for handling and securing the model are clearly defined and assigned to trained personnel\\\hline
    \hline
    \multicolumn{2}{|c|}{\textbf{M.02 Model storage}}\\\hline
        Not secure & model data is not secured by information security controls -- confidentiality, integrity, and availability can be compromised\\\hline
        Secure & confidentiality and integrity is protected by security controls, but availability can be affected\\\hline
        Secure and guaranteed & like Secure, but with added measures to protect availability to support business continuity\\\hline
    \hline
    \multicolumn{2}{|c|}{\textbf{M.03 Model robustness}}\\\hline
        Not implemented & adversarial attacks are not taken into account when the models are constructed\\\hline
        Partially implemented & security-critical models are trained to be robust against pre-defined set of adversarial attacks\\\hline
        Fully implemented & all the model are trained to be robust against potential adversarial attacks\\\hline
    \hline
    \multicolumn{2}{|c|}{\textbf{M.04 Change control}}\\\hline
        Undefined & there is no change control in place -- when new model is created, it replaces the old one\\\hline
        Semi-defined & there is a repository of older models, but no formalized way to rollback the changes\\\hline
        Defined & there is a formal process for change control in place\\\hline
    \hline
    \multicolumn{2}{|c|}{\textbf{M.05 Computational redundancy}}\\\hline
        Not implemented & there is no redundant computation taking place, all the results are obtained from a single model\\\hline
        Dual redundancy & there are two different models in place, if the result is not the same, it is not used\\\hline
        Multiple redundancy & there are multiple models in place, the result is obtained by majority voting\\\hline
    \hline
    \multicolumn{2}{|c|}{\textbf{M.06 Model deployment guarantees}}\\\hline
        Third-party deployment & no knowledge of the deployment environment, no security guarantees \\\hline
        Client deployment & partial knowledge of the deployment environment, partial security guarantees\\\hline
        In-house deployment & full knowledge of the environment, full security guarantees\\\hline
    \hline
    \multicolumn{2}{|c|}{\textbf{M.07 Runtime validation}}\\\hline
        Undefined & there are no runtime validation processes in place\\\hline
        Semi-defined & runtime validation is performed, but there is no formal process on how to interpret the results\\\hline
        Defined & there is a formal process to guide the runtime validation and to interpret the results\\\hline
\end{supertabular}
\end{center}

\section{Execution Environment Evaluation Factors}
\label{app:sec_ef_exec}

\begin{center}\footnotesize
\tablecaption{Execution environment evaluation factors defined in the framework.} 
 \label{app:ef_exec_eval}
\begin{supertabular}{|p{3cm}|p{9.5cm}|}\hline 
    \multicolumn{2}{|c|}{\textbf{E.01 System patches and versioning}}\\\hline
        Not implemented & there are no formal processes for patching and versioning of systems, updates are implemented ad-hoc\\\hline
        Partially implemented & critical patches are implemented, but there is no versioning that could revert the potentially unwanted changes\\\hline
        Fully implemented & processes are set to keep the systems to date with the latest patches and versioning is implemented to be able to revert the changes in case it is needed\\\hline
    \hline
    \multicolumn{2}{|c|}{\textbf{E.02 Software integrity}}\\\hline
        Not implemented & integrity checking mechanisms for software are not implemented  \\\hline
        Partially implemented & critical software components are periodically subjected to integrity checks\\\hline
        Fully implemented & all software components of critical systems are periodically subjected to integrity checks\\\hline
    \hline
    \multicolumn{2}{|c|}{\textbf{E.03 Supply chain security}}\\\hline
        Undefined & there are no processes in place that would mitigate potential security incidents related to supply chain\\\hline
        Semi-defined & some processes for supply chain security are in place, but they are not formally defined\\\hline
        Defined & processes are in place to handle security related to supply chain\\\hline
        \hline
    \multicolumn{2}{|c|}{\textbf{E.04 Backups}}\\\hline
        Not implemented & procedures for information system backups are not in place \\\hline
        Partially implemented & procedures for backups of  critical information systems are defined\\\hline
        Fully implemented & formal procedures for backups for all information systems are in place\\\hline
        \hline
    \multicolumn{2}{|c|}{\textbf{E.05 Maintenance}}\\\hline
        Not implemented & procedures for maintenance and repair of information systems are not in place \\\hline
        Partially implemented & procedures for maintenance and repair of critical information systems are defined\\\hline
        Fully implemented & formal procedures for maintenance and repair for all information systems are in place\\\hline
        \hline
    \multicolumn{2}{|c|}{\textbf{E.06 Network monitoring}}\\\hline
        Not implemented & monitoring of computer networks for anomalies and incidents is not implemented \\\hline
        Partially implemented & some parts of the computer networks are monitored for anomalies and incidents\\\hline
        Fully implemented & formal procedures for monitoring computer networks for anomalies and incidents are defined\\\hline
        \hline
    \multicolumn{2}{|c|}{\textbf{E.07 Deployment validation}}\\\hline
        Not implemented & ML-based systems are not validated at the place of deployment \\\hline
        Partially implemented & deployment validation of ML-based systems is performed, but there are no formal procedures\\\hline
        Fully implemented & formal procedures for deployment validation of ML-based systems are in place\\\hline
\end{supertabular}
\end{center}

\section{Security Controls Evaluation Factors}
\label{app:sec_ef_controls}
\begin{center}\footnotesize
\tablecaption{Security controls evaluation factors defined in the framework.} 
 \label{app:ef_sec_eval}
\begin{supertabular}{|p{3cm}|p{9.5cm}|}
    \hline
    \multicolumn{2}{|c|}{\textbf{S.01 Cybersecurity roles and responsibilities}}\\\hline
        Undefined & there are no cybersecurity roles defined within the organization\\\hline
        Semi-defined & cybersecurity roles are present in the organization but their responsibilities are not clearly defined\\\hline
        Defined & cybersecurity roles and responsibilities within the organization are clearly defined\\\hline
    \hline
    \multicolumn{2}{|c|}{\textbf{S.02 Accountability}}\\\hline
        Not implemented & there are no processes in place related to accountability\\\hline
        Partially implemented & actions are logged on some devices, but the accountability is not strictly enforced\\\hline
        Fully implemented & all the processes that are required for accountability are clearly defined\\\hline
    \hline
    \multicolumn{2}{|c|}{\textbf{S.03 Security awareness and training}}\\\hline
        Not implemented & there is no security awareness in the organization and no security-related trainings \\\hline
        Partially implemented & security-critical roles are trained to obey best practices, but security awareness is not present throughout the entire organization\\\hline
        Fully implemented & security awareness is present in the entire organization and regular trainings are conducted to keep employees up to date with the latest practices\\\hline
    \hline
    \multicolumn{2}{|c|}{\textbf{S.04 Identity management and access control}}\\\hline\hline
        Not implemented & there are no processes in place related to identity management and access control\\\hline
        Partially implemented & some processes for identity management are implemented but access control is not clearly defined w.r.t. organizational roles\\\hline
        Fully implemented & identity management processes are clearly defined to provide procedures for user enrollment, authorization and authentication\\\hline
      \hline
    \multicolumn{2}{|c|}{\textbf{S.05 Business continuity and disaster recovery}}\\\hline
        Not implemented & there is no business continuity implemented in the organization\\\hline
        Partially implemented & some processes are in place, but they are not formally defined -- timeline for disaster recovery is not clear\\\hline
        Fully implemented & processes are defined to handle unexpected events that can disturb business continuity\\\hline  
    \hline
    \multicolumn{2}{|c|}{\textbf{S.06 Incident reporting}}\\\hline
        Not implemented & procedures for incident reporting are not implemented \\\hline
        Partially implemented & incident reporting is done to some extend, but there are no formal procedures in place\\\hline
        Fully implemented & formal procedures for incident reporting are defined\\\hline
    \hline
    \multicolumn{2}{|c|}{\textbf{S.07 Incident response}}\\\hline
        Not implemented & procedures for incident response are not implemented \\\hline
        Partially implemented & incident response is done to some extend, but there are no formal procedures in place\\\hline
        Fully implemented & formal procedures for incident response are defined\\
    \hline
    \multicolumn{2}{|c|}{\textbf{S.08 Forensics}}\\\hline
        Not implemented & processes for forensics investigation after an incident are not implemented \\\hline
        Partially implemented & processes for forensics investigation after a major incident are in place\\\hline
        Fully implemented & formal processes for forensics investigation after an incident are in place\\\hline
    \hline
    \multicolumn{2}{|c|}{\textbf{S.09 Incident mitigation strategy}}\\\hline
        Not implemented & strategies to mitigate incidents are not revised after an incident \\\hline
        Partially implemented & strategies to mitigate incidents are revised after a major incident\\\hline
        Fully implemented & there is a formal procedure on incident mitigation strategy -- incidents are revisited after an incident and updated to incorporate lessons learned\\\hline
    \hline
    \multicolumn{2}{|c|}{\textbf{S.10 Communication of recovery activities}}\\\hline
        Not implemented &  there is no process for communicating of recovery activities to relevant parties \\\hline
        Partially implemented & recovery activities after major incident are communicated to relevant parties\\\hline
        Fully implemented & formal procedures for communicating of recovery activities to relevant parties are in place \\\hline
\end{supertabular}
\end{center}

\end{document}